\documentclass[nohyper]{JHEP3}
\usepackage{amssymb,amsxtra}
\usepackage{epsfig}              
\usepackage{latexsym}
\usepackage{bm}
\usepackage[T1]{fontenc}

\newcommand{\nn}{\nonumber}
\newcommand{\N}{\mathbb}
\newcommand{\mc}{\mathcal}

\newcommand{\half}{{\textstyle\frac{1}{2}}}
\newcommand{\fourth}{{\textstyle\frac{1}{4}}}

\renewcommand{\Re}{{\rm Re\thinspace}}
\renewcommand{\Im}{{\rm Im\thinspace}}

\usepackage{amsmath}

\title{Symmetry and Mass Degeneration in Multi-Higgs-Doublet Models}
\author{K. Olaussen,\\
Department of Physics,
NTNU, N-7491 Trondheim, Norway\\
E-mail: \email{Kare.Olaussen@ntnu.no}}
\author{P. Osland\\ 
Department of Physics and Technology, 
University of Bergen, Postboks 7803, N-5020 Bergen, Norway\\ 
E-mail: \email{Per.Osland@ift.uib.no}}
\author{M. Aa. Solberg\\
Department of Physics,
NTNU, N-7491 Trondheim, Norway\\
E-mail: \email{marius@tf.phys.ntnu.no}}

\abstract{We investigate possible symmetry properties of the scalar sector of
Multi-Higgs-Doublet Models, and, to some extent, the generalization of such
models to gauge groups other than $SU(2)_L\times U(1)_Y$. In models 
with $C$ (charge conjugation) invariance, and where certain quartic terms
are not present, the scalar potential is invariant under a group larger than the
gauge group, $O(4)$ when the Higgs fields are doublets.  If the Higgs fields
develop aligned vacuum expectation values, this symmetry will break to an
$O(3)$ subgroup, which in general is further broken by
loop corrections involving the gauge bosons.  Assuming such corrections are small, the physical properties of the Higgs sector will
approximately organize into representations
of $SO(3)$.  If the vacuum expectation values of the Higgs
fields are aligned in the direction of the $C$ even fields, the mass spectra
of the charged and $C$ odd sectors will be degenerate. 
Moreover, if the Higgs fields develop a pair
of non-aligned vacuum expectation values, so that
the charge conjugation symmetry is spontaneously broken
(but not the $U(1)$ electromagnetic gauge invariance), a
pair of light charged pseudo-Goldstone bosons will appear.} 


\begin{document}

\section{Introduction}
\label{sec:Introduction}

Future discoveries at the LHC may reveal a sector of scalar particles that is
much richer than that of the Standard Model (SM). Some of the scalars may be
responsible for generating the masses of fermions and the electroweak bosons
\cite{Englert:1964et}, whereas others may be responsible for the dark matter
\cite{Silveira:1985rk,Holz:2001cb,Ma:2006km,Barbieri:2006dq,Patt:2006fw,
Porto:2007ed,Grzadkowski:2009bt}. 
Another interesting aspect of these theories is that CP violation is naturally accommodated, including its spontaneous breaking \cite{Lee:1973iz}.

It is natural to classify such scalars according to their properties under the
$SU(2)$ associated with the electroweak sector of the Standard Model.  In
order to be compatible with electroweak precision data, one usually considers
only $SU(2)$ doublets and singlets. Even these representations are severely
constrained by the data
\cite{Gunion:2002in,WahabElKaffas:2007xd,Grimus:2007if,Haller:2008tn}.

Going beyond one or two doublets \cite{Gunion:1989we,Davidson:2005cw,Barroso:2007rr}, one
immediately has to face models having a large number of parameters.  The
structure of such potentials has been studied in \cite{Nishi:2007nh}.  

Different doublets could be distinguished via their couplings to fermion
fields. This idea is exploited in the so-called Model~II version of the
two-Higgs-doublet model (2HDM) \cite{Gunion:1989we}, where one doublet couples
to up-type fermions, and the other couples to down-type fermions.  Another
version of this idea is the one considered in Ref.~\cite{Porto:2007ed}, where
each fermion or each family has its own Higgs field.

We shall here consider instead the case when the different doublets can not be
distinguished (since we are not considering couplings to the fermions). An introduction of Yukawa couplings would naturally have broken the symmetry among the different doublets \cite{Haber:1992py}. Thus, we shall here study the symmetries of models with $N$ doublets---it turns out that by setting a certain $SO(4)$-violating parameter $\lambda^{(3)}$ to zero and assuming vacuum alignment, the spectrum
simplifies considerably. In particular, a certain ``custodial'' $SO(3)$
symmetry \cite{Sikivie:1980hm,Willenbrock:2004hu} leads to a degeneracy between the mass matrix of the $C$ odd (or equivalently, $CP$ odd) and the
charged Higgs bosons.

This possibility of a symmetry group of the scalar polynomial which is larger than required by gauge invariance was pointed out
by Weinberg \cite{Weinberg:1972fn} many years ago.  In the theories considered
the extra symmetry was assumed to be a symmetry of any quartic
(i.e. renormalizable) potential of the scalar sector.

The Standard Model with its single Higgs doublet is an example of a theory
were the most general scalar potential has an extra $O(4)$ symmetry not
contained in the $SU(2)\times U(1)$ gauge symmetry.  
An extension of the SM with an extra Higgs doublet (as for instance required if we want to
introduce supersymmetry) or more, and with $C$ symmetry, also has
a quadratic potential which is automatically $O(4)$ symmetric.
This is sufficient to enforce $O(4)$ mass relations up to perturbative corrections in the parameter $g'$. (Hence these corrections involve the gauge bosons $Z$ and $\gamma$.)  For the quartic potential, the extra $O(4)$ symmetry is broken by the parameter(s)
$\lambda^{(3)}$, cf. eq.~\eqref{E:potNHDMCPinv}. 
Standard renormalizability instructs us to include into the
Lagrangian, {\em all\/} terms allowed by the $SU(2)\times U(1)$ symmetry,
hence it may be inconsistent to leave out terms proportional to $\lambda^{(3)}$. The presence of
such parameters will in general lead to an order $\lambda^{(3)}$ tree-level
breaking of the additional symmetry. This is in principle not different from having the symmetry broken by
loop corrections. It becomes in any case a question of the magnitude of the
perturbation. 

We present a detailed classification of the possible terms in the potential, discussing charge conjugation and how the familiar custodial $O(4)$ symmetry of the SM potential generalizes to an $O(4)$ (or, more generally, $O(2k)$) symmetry for certain terms of the potential in the NHDM. The maximal $O(2k)$ symmetric potential turns out to be a more constrained potential than the maximal charge-invariant one. The kinetic terms are in general subject to an independent classification, depending on the $U(1)$ hypercharge coupling $g'$. We identify, in certain situations, a charged pair whose mass vanishes with $g'$.

The paper is organized as follows. In section~\ref{sec:potential} we discuss the general potential and classify the corresponding Lagrangian, including the kinetic terms. In section~\ref{sec:ssb} we discuss spontaneous symmetry breakdown, and in section~\ref{sec:conclusion} we conclude. A couple of more matematical discussions are delegated to appendices.
\section{The NHDM potential and Lagrangian}
\label{sec:potential}
We define the $N$-Higgs-doublet model, abbreviated NHDM, to be a system
of $N$ two-com\-po\-nent complex scalar fields
$\Phi_1, \Phi_2, \ldots, \Phi_N$, each with the same transformation
property under $SU(2)_L\times U(1)_Y$ as the Higgs field of
the Standard Model, and with dynamics defined by the
Lagrangian density
\begin{align}\label{E:fullNHDMpot}
 \mc{L}(x)&= \sum_m\; [D^\mu \Phi_m(x)]^\dag [D_\mu \Phi_m(x)] 
 - V(\Phi_1, \Phi_2,\ldots, \Phi_N),
\end{align}
where $V(\Phi_1, \Phi_2,\ldots, \Phi_N)$ is a potential that---in its most
general form---is given by \eqref{E:potNHDM} below.
The \emph{covariant derivative} $D^\mu$ is defined as
\begin{equation}\label{E:CovDer1}
 D^{\mu} = \partial^\mu + ig T_iW_i^\mu + ig'YB^\mu,
\end{equation}
where $W_i^\mu$ and $B^\mu$ are the $SU(2)_L$ and $U(1)_Y$ gauge fields,
respectively, and  $T_i = \frac{1}{2}\sigma_i$ are the generators of $SU(2)$
with $\sigma_1, \sigma_2, \sigma_3$ the Pauli matrices. Thus, our Higgs fields
are labeled by two indices: The {\em row index\/} running from 1 to $N$ is
often written out explicitly as above, and an often hidden
{\em group index\/} acted on by the gauge group. The latter are acted on
by the matrices $T_i$ in \eqref{E:CovDer1} (whose indices are also hidden). 
When written explicitly we shall use Greek letters
from the beginning of the alphabet.

To write the most general gauge-invariant potential in a renormalizable NHDM
in a compact way, we introduce a set of linearly independent\footnote{
There are no linear relations between the operators in \eqref{E:opNHDM}.
However, they are algebraically dependent when $N\ge3$, being restricted by
$(N-2)^2$ polynomial equations of 8'th order in the fields.}
hermitian operators
invariant under local $SU(2)_L\times U(1)_Y$ transformations (this is a generalization of the approach for the 2HDM in
\cite{Diaz:2002tp}):
\begin{align}\label{E:opNHDM}
 \widehat{A}_m   &= \phantom{-}\Phi_m^\dag \Phi_m, \nonumber \\
 \widehat{B}_{mn} &= \phantom{-}\half(\Phi_m^\dag \Phi_{n} + \Phi_{n}^\dag
 \Phi_{m}) =\text{Re}(\Phi_m^\dag \Phi_{n}) \equiv \widehat{B}_{a},\\
 \widehat{C}_{mn} &= -{\textstyle\frac{i}{2}}
 (\Phi_m^\dag \Phi_{n} - \Phi_{n}^\dag
 \Phi_{m}) =\text{Im}(\Phi_m^\dag \Phi_{n}) \equiv \widehat{C}_{a}.\nonumber
\end{align}
Due to (anti-)symmetry under interchange of $m$ and $n$ we may impose the
restriction that $1\le m< n\le N$, and introduce indices $a,\,b,\ldots$
labeling such pairs. An explicit invertible encoding is
\begin{equation}\label{E:enc} 
 1\le a=a(m,n)=m+\frac{1}{2}(n-2)(n-1)\le \frac{1}{2}N(N-1)\equiv {\cal N}.
\end{equation}
We let $m(a),n(a)$ denote the inverse of this encoding.  We will use the
summation convention that repeated indices from the beginning of the alphabet
are summed from 1 to ${\cal N}$, and repeated indices from the middle of the
alphabet are summed from 1 to $N$.  The most general potential in the NHDM
thus becomes a linear combination of all different quadratic and quartic
factors in the $\Phi_{m}$ (and $\Phi^\dag_{m}$) which can be formed from
$\widehat{A}_{m}$, $\widehat{B}_{a}$ and $\widehat{C}_{a}$:\footnote{Since
$\left(\Phi^\dag_k\sigma^j\Phi_\ell\right)
\left(\Phi^\dag_m\sigma^j\Phi_n\right)
=-\left(\Phi^\dag_k\Phi_\ell\right)\left(\Phi^\dag_m\Phi_n\right) +
2\left(\Phi^\dag_k\Phi_n\right)\left(\Phi^\dag_m\Phi_\ell\right)$, other
quartic invariants may be expressed by those chosen.}
\begin{align}\label{E:potNHDM}
 V_\text{g}(\Phi_1, \ldots,\Phi_N) &= \mu_m^{(1)}\widehat{A}_m
 +\mu_{a}^{(2)}\widehat{B}_{a} + \mu_{a}^{(3)}\widehat{C}_{a}
 + \lambda_{mn}^{(1)} \widehat{A}_m\widehat{A}_n +
 \lambda_{ab}^{(2)}\widehat{B}_{a}\widehat{B}_{b}\nonumber \\ 
 &+ \lambda_{ab}^{(3)} \widehat{C}_{a}\widehat{C}_{b} 
 + \lambda_{ma}^{(4)} \widehat{A}_{m}\widehat{B}_{a} 
 + \lambda_{ma}^{(5)}  \widehat{A}_{m}\widehat{C}_{a}
 + \lambda_{ab}^{(6)} \widehat{B}_{a} \widehat{C}_{b},
\end{align}
where the ``g'' in $V_\text{g}$ denotes ``general''. To avoid double counting
we introduce the restriction $m\le n$ in the term involving
$\lambda_{mn}^{(1)}$, and the restriction $a\le b$ in the terms involving
$\lambda_{ab}^{(2)}$, $\lambda_{ab}^{(3)}$ and $\lambda_{ab}^{(6)}$.  We will
not consider terms of degree higher than four, because these would destroy the
renormalizability of the model \cite{Cheng&Li}.
From the hermiticity of the potential
$V_\text{g}$ all parameters $\mu$ and $\lambda$ in the expansion
\eqref{E:potNHDM} must be real. Thus the number of free real parameters in
~\eqref{E:potNHDM} is
\begin{equation}
  N_\text{tot}= N+2{\cal N} + \half N(N+1)+ {\cal N}({\cal N}+1)
  + 2 N{\cal N}+{\cal N}^2 = \half N^2(N^2+3),
\end{equation}
which for $N=1$ gives us the 2 parameters of the Standard Model
($\mu^2$ and $\lambda$). $N=2$ gives us the usual 14 parameters for the
2HDM. There are 54 parameters for $N=3$ and $152$ parameters for $N=4$.

This counting does not take into account the fact that we may make $SU(N)$ row
transformations on the fields $\Phi_m$ to eliminate some terms in
\eqref{E:potNHDM}. One possible choice is to transform the quadratic terms
into a diagonal form, i.e. so that $\mu^{(2)}_a = \mu^{(3)}_a=0$.  This in
general leaves a matrix of $N-1$ independent diagonal phase transformations
(such that the determinant is unity). We may for instance use it to transform
all $\lambda_{1a}^{(5)}$ with $n(a)=m(a)+1$ to zero. This reduces the number
of parameters by $N^2-1$, i.e. to
${N}'_\text{tot}=\half\left(N^4+N^2+2\right)$, yielding 11 parameters for
$N=2$ (in agreement with Davidson and Haber \cite{Davidson:2005cw}), 46
parameters for $N=3$, and 137 parameters for $N=4$.

\subsection{The most general $C$-invariant NHDM-potential} 

The charge conjugation operator $C$ is a linear operator, multiplicative in the fields, which leaves
complex constants unaltered, but maps fields onto their hermitian conjugate transposed;
$C(z\Phi_m)=z\Phi_m^{\dagger \, T}$, where $z$ is a complex
number.\footnote{This definition assumes that we for some reasons have decided
on a decomposition of all fields into their real and imaginary parts. It is
not invariant under complex transformations of the fields, see e.g.\
\cite{BrancoRebeloSilva-Marcos2005}.}  Then 
$C(\widehat{C}_{a})=-\widehat{C}_{a}$, in contrast to 
$C(\widehat{A}_{m})=\widehat{A}_{m}$, and 
$C(\widehat{B}_{a})=\widehat{B}_{a}$.  We obtain a $C$-invariant
potential by leaving out all terms which are odd in $\widehat{C}_{a}$, i.e.,
terms involving $\mu_{a}^{(3)}$, $\lambda_{ma}^{(5)}$, and
$\lambda_{ab}^{(6)}$. There are ${\cal N} + N {\cal N} + {\cal N}^2 = \fourth
N(N-1)(N^2+N+2)$ such terms, leaving
\begin{equation}
 N_C = \frac{1}{4}N(N^3+5N+2)
\end{equation}
free parameters for the general renormalizable $C$-invariant potential,
\begin{align}\label{E:potNHDMCPinv}
 V_C(\Phi_1, \ldots,\Phi_N) &= \mu_m^{(1)}\widehat{A}_m
 +\mu_{a}^{(2)}\widehat{B}_{a}
 + \lambda_{mn}^{(1)} \widehat{A}_m\widehat{A}_n
 + \lambda_{ab}^{(2)}\widehat{B}_{a}\widehat{B}_{b}\nonumber\\
 &+ \lambda_{ab}^{(3)} \widehat{C}_{a}\widehat{C}_{b} 
 + \lambda_{ma}^{(4)} \widehat{A}_{m}\widehat{B}_{a}.
\end{align}
For $N=1$ we get the usual 2 parameters of the standard model, the Higgs
potential of which is automatically $C$-invariant.  For $N=2$ we get
the usual (see e.g., \cite{Diaz:2002tp}) 10 parameters. For $N=3$ we get 33
parameters, and for $N=4$ we get 86 parameters.

This counting does not take into account that we may make $O(N)$
transformations on the row of fields $\Phi_m$ to eliminate
some terms in \eqref{E:potNHDMCPinv}. A natural choice is to transform
the quadratic terms to diagonal form,
i.e.\ so that $\mu_{a}^{(2)}=0$. This reduces the number of parameters
by $\half N(N-1)$, i.e.\ to $N'_C=\fourth N(N^3 + 3N +4)$.
This gives 9 parameters for $N=2$ (in agreement with \cite{Barroso:2007rr}),
30 parameters for $N=3$, and 80 parameters for $N=4$.

The difference $N_\text{phases}=N'_\text{tot}-N'_C =
\frac{1}{4}N^2(N^2-1)+1-N$ counts the number of genuine $C$-violating
parameters in $V_\text{g}$ (in agreement with Branco {\em et
al. }\cite{BrancoRebeloSilva-Marcos2005}).

\subsection{Symmetries of $\bm{\widehat{A}}$, $\bm{\widehat{B}}$, 
$\bm{\widehat{C}}$ and $\bm{\widehat{C}^2}$}
\label{sec:SymmetriesOfABC}

For generality we here consider $k$ (rather than 2)-component fields, i.e.\
with $SU(k)\times U(1)$ as gauge group.  To make it simpler to explore all
possible symmetries we express the field $\Phi_m$ in terms of its independent
real (hermitian) components, \( \Phi_m = \Psi_m + \text{i} \Theta_m \). Define
$2k\times 2k$ matrices
\begin{equation} \label{E:calI-calJ}
{\cal I}=\left(\begin{array}{rr}I_k&0_k\\0_k&I_k\end{array}\right), \quad
{\cal J}=\left(\begin{array}{rr}0_k&I_k\\-I_k&0_k\end{array}\right),
\end{equation}
where the subscript $k$ indicates the linear dimension of the submatrix
involved.  The complex scalar product between two fields $\Phi_m$ and
$\Phi_n$, invariant under unitary ($U(k)$) transformations, can be expressed
in terms of two real bilinear forms\footnote{$\left\langle \Phi_m, \Phi_n \right\rangle = \Phi_m^\ast \cdot \Phi_n ={\widehat B}_{mn} +i {\widehat C}_{mn}$. }
\begin{subequations} \label{E:Bhat-Chat}
\begin{align} 
  &\text{Re}(\Phi_m^\dag \Phi_{n}) 
= {\widehat B}_{mn} = \left(\Psi_m^T,\Theta_m^T \right)
  {\cal I}
   \left(\begin{array}{c}\Psi_n\\\Theta_n\end{array}\right)
=  \Psi_m^T \Psi_n + \Theta_m^T \Theta_n,\\
  &\text{Im}(\Phi_m^\dag \Phi_{n}) 
= {\widehat C}_{mn} = \left(\Psi_m^T,\Theta_m^T \right)
  {\cal J}
   \left(\begin{array}{c}\Psi_n\\\Theta_n\end{array}\right) 
= \Psi_m^T \Theta_n - \Theta_m^T \Psi_n.
\end{align}
\end{subequations}
The first is the Euclidean dot product between $2k$-component real vectors,
the second is the Poisson bracket (symplectic form) of the same quantities
viewed as coordinates and momenta of $2k$-dimensional phase space.
The quantities in \eqref{E:Bhat-Chat} are individually invariant under
transformation groups larger than $U(k)$.  The first form $\widehat{B}$ (with $\widehat{A}$ as a special case) is invariant under
the $O(2k)$ group of real orthogonal transformations,
\begin{equation}\label{E:orthogonal}
 \left(\begin{array}{c}\Psi_n\\
   \Theta_n\end{array}\right) \rightarrow O
   \left(\begin{array}{c}\Psi_n\\
     \Theta_n\end{array}\right),\quad
     O^T O = {\cal I},
\end{equation}
the second form $\widehat{C}$ is invariant under the $Sp(k,R)$ group of real symplectic
transformations,
\begin{equation}\label{E:symplectic}
 \left(\begin{array}{c}\Psi_n\\
   \Theta_n\end{array}\right) \rightarrow S
   \left(\begin{array}{c}\Psi_n\\
     \Theta_n\end{array}\right),\quad
     S^T {\cal J} S = {\cal J}.
\end{equation}
In this formulation the charge conjugation operator $C$ discussed above
can be represented as a particular $O(2k)$ transformation when acting on the
fields $\Psi_n$ and $\Theta_n$
\begin{equation}\label{E:calC}
  C = 
  \left(\begin{array}{rr}
    I_k&0_k\\0_k&-I_k
  \end{array}\right).
\end{equation}
Considering infinitesimal transformations, $O={\cal I}+\epsilon L + {\cal
O}(\epsilon^2)$, $S={\cal I}+\epsilon M + {\cal O}(\epsilon^2)$, the conditions
\eqref{E:orthogonal} and \eqref{E:symplectic} become
\begin{equation}\label{E:infinitesimalEq}
 L^T {\cal I} + {\cal I} L = 0_{2k},\quad M^T {\cal J} + {\cal J} M = 0_{2k}.
\end{equation}
Thus $L$ must be a $2k\times 2k$ antisymmetric real matrix; there is
a set (Lie algebra) $so(2k)$ of $2k^2-k$ linearly independent such matrices.
Writing out the condition for $M$ in
terms of $k\times k$ submatrices we find that it must have the form
\begin{equation}\label{E:Mspk}
 M = \left(\begin{array}{cc}A&B\\
   C&-A^T\end{array}\right),\quad B=B^T,\;
 C=C^T.
\end{equation}
There is a set $sp(k)$ of $k^2 + k(k+1) = 2k^2+k$ linearly independent such
matrices. The infinitesimal transformations of the original $U(k)$ are the
intersection of the sets $so(2k)$  and $sp(k)$. I.e., the matrices
of the form 
{\scriptsize$\begin{pmatrix}A&B\\-B&A\end{pmatrix}$},
with $A=-A^T$ and
$B^T=B$.  There are $\half k(k-1) + \half k(k+1) = k^2$ such linearly
independent matrices.

\paragraph{The symmetries of $\widehat{C}^2$}
\label{sec:TheSymmetriesOfWidehatC2}
The form $\widehat{C}^2$ (or more precisely $\widehat{C}_{mn} \widehat{C}_{m'n'}$) 
has a bigger symmetry group than the form $\widehat{C}$. Still, we will see that such operators (forms) will violate the full $O(4)$ symmetry we can assign the rest of the Lagrangian. In analogy with \eqref{E:symplectic}, $\widehat{C}^2$
symmetries are given by
\begin{align}\label{E:symplectic2}
	\left(\begin{array}{c}\Psi_n\\
   \Theta_n\end{array}\right) \rightarrow S
   \left(\begin{array}{c}\Psi_n\\
     \Theta_n\end{array}\right),\quad
     S^T {\cal J} S = \pm {\cal J},
\end{align}
which can be collected in a set 
\begin{align}\label{Def:Pk}
	P(k,\N{R})=\{ S\in GL_{2k}(\N{R}) | S^T {\cal J} S = \pm {\cal J} \},
\end{align}
which we in Appendix \ref{sec:TheSymmetryGroupOfWidehatC2} show is a Lie group.

The component
\begin{align}
	P^-(k,\N{R})=\{ S\in GL_{2k}(\N{R}) | S^T {\cal J} S = - {\cal J} \},
\end{align}
  consists of matrices with determinant 
\begin{align}\label{E:Det(-1)^k}
	\det(P^-(k,\N{R})) =(-1)^k,
\end{align}
   as shown in appendix \ref{sec:TheSymmetryGroupOfWidehatC2}.
The group $P(k,\N{R})=Sp(k,\N{R})\cup P^-(k,\N{R})$ will have the same Lie algebra as $Sp(k,\N{R})$, since the new component $P^-(k,\N{R})$ is not connected with the identity. This is manifested by the equation corresponding to eq.~\eqref{E:infinitesimalEq},
\begin{equation}\label{E:condC2}
{\cal J}+ \epsilon (M^T {\cal J} + {\cal J} M) = \pm {\cal J}
\end{equation}
not having any solution for the case of a $-{\cal J}$ on the right side, see appendix \ref{sec:TheSymmetryGroupOfWidehatC2} for a proof.

\paragraph{The most general $O(2k)$-symmetric potential}
 We can conclude that the most general $O(2k)$-invariant potential can be written
 \begin{align}\label{E:potNHDMO2kinv}
 V_{O(2k)}(\Phi_1, \ldots,\Phi_N) &= \mu_m^{(1)}\widehat{A}_m
 +\mu_{a}^{(2)}\widehat{B}_{a}
 + \lambda_{mn}^{(1)} \widehat{A}_m\widehat{A}_n \nonumber\\
 &+ \lambda_{ab}^{(2)}\widehat{B}_{a}\widehat{B}_{b}
 + \lambda_{ma}^{(4)} \widehat{A}_{m}\widehat{B}_{a},
\end{align}
since we have seen that operators not containing any factor $\widehat{C}$ are $O(2k)$-invariant.
We obtain an $O(2k)$-invariant
potential by leaving out terms proportional
to $\lambda_{ab}^{(3)}$ from the $C$ invariant potential $V_C$ \eqref{E:potNHDMCPinv}. The number of terms in $V_{O(2k)}$ is then $N_{O(2k)}=N_C- {\cal N}^2  = \fourth N(N^3+5N+2)- \fourth N^2 (N-1)^2$, giving
\begin{equation}
 N_{O(2k)} = \frac{1}{2}N(N+1)^2
\end{equation}
free parameters for the general renormalizable $O(2k)$-invariant potential $V_{O(2k)}$.
For $N=1$ we get the usual 2 parameters of the standard model, the Higgs
potential being automatically $O(2k)$-invariant.  For $N=2$ we get
the usual 9 parameters, one parameter less than for the $C$-invariant potential. For $N=3$ we get 24
parameters, and for $N=4$ we get 50 parameters.

This counting does again not take into account that we may make $O(N)$
transformations on the row of fields $\Phi_m$ to eliminate
some terms in \eqref{E:potNHDMO2kinv}. We may transform
the quadratic terms to diagonal form,
so that $\mu_{a}^{(2)}=0$. This reduces the number of parameters
by $\half N(N-1)$, i.e.\ to $N'_{O(2k)}=\half N(N^2 + N +2)$.
This gives 8 parameters for $N=2$,
21 parameters for $N=3$, and 44 parameters for $N=4$.
\subsection{Symmetries of the NHDM potential}
\label{sec:SymmetriesOfTheNHDMPotentials}

Since the NHDM-potential $V_{\text{g}}$ is constructed from the invariants
\eqref{E:opNHDM} the symmetries of the latter are reflected
in the symmetries of the former, but in a manner depending on
details of the construction:
\begin{enumerate}
\item If $V_{\text{g}}$ depends only on the ${\widehat A}_m$'s, i.e. if only
 the parameters $\mu_m^{(1)}$ and $\lambda_{mn}^{(1)}$ are nonzero, then the
 symmetry group of $V_{\text{g}}$ is at least\footnote{It could possibly be
 larger, since there might be additional row symmetries transforming
 fields $\Phi_m$ with different $m$ into each other; such symmetries would
 require special relations among the parameters $\mu_m^{(1)}$ and
 $\lambda_{mn}^{(1)}$.}  $\bigotimes_{m=1}^N O(2k)$, since we can make
 independent transformations on each $\Phi_m$.

\item If $V_{\text{g}}$ depends only on the ${\widehat A}_m$'s and the
 $\widehat{B}_a$'s, i.e. for a $C$-invariant theory
 \eqref{E:potNHDMCPinv}, where in addition the parameters
 $\lambda_{ab}^{(3)}=0$, then the symmetry group of $V_{\text{g}}$ is at
 least $O(2k)$. It may contain several such factors if some of the parameters
 $\mu_{a}^{(2)}$ and $\lambda_{ab}^{(2)}$ vanish. To analyze this we
 partition the $\Phi_m$'s into sets: If a parameter $\mu_{a}^{(2)}$ is
 nonzero, then the fields $\Phi_{m(a)}$ and $\Phi_{n(a)}$ belong to the same
 set, with $m(a)$ and $n(a)$ denoting that $m$ and $n$ are contained in
 $a$. If a parameter $\lambda_{ab}^{(2)}$ is nonzero, then the fields
 $\Phi_{m(a)}$ and $\Phi_{n(a)}$ belong to the same set, and the fields
 $\Phi_{m(b)}$ and $\Phi_{n(b)}$ belong to the same set.  With this
 partitioning into a maximal number of sets we may make one independent
 $O(2k)$ transformation for each set.

\item If $V_{\text{g}}$ depends only on the ${\widehat C}_a$'s, i.e. with only
 the parameters $\mu_{a}^{(3)}$ and $\lambda_{ab}^{(3)}$ being nonzero, then
 the symmetry group of $V_{\text{g}}$ is at least $Sp(k,\N{R})$. If we (in the
 same manner as above) can partition the fields into several sets, then we may
 make independent $Sp(k,\N{R})$ transformations on fields belonging to different
 sets. However, since the additional symmetries in this case fail to be
 symmetries of even the zero'th order kinetic term \eqref{E:kinetic0}, their
 significance is uncertain.

\item With all parameters arbitrary the symmetry group of $V_{\text{g}}$ is
 just the original $SU(k)\times U(1)$ gauge symmetry.

\end{enumerate}

\noindent
In this work we will pay special attention to the second scenario, with $k=2$.

\subsection{Symmetries of the kinetic terms}
\label{sec:SymmetriesOfTheKineticTerms}
We now turn to the (global) symmetries of the 
kinetic terms of the Lagrangian,
\begin{align}
K= \sum_{n=1}^N	[(\partial^\mu + G^\mu)\Phi_n(x)]^\dag [(\partial_\mu + G_\mu)\Phi_n(x)],
\end{align}
with
\begin{align}\label{def:G}
	G^\mu= ig T_iW_i^\mu + ig'YB^\mu.
\end{align}
Let $K_i$ denote the terms of the $i$'th order in the gauge fields.

Consider the transformation of the kinetic terms linear in the gauge fields, $K_1$, under the map $\rho$ defined in appendix \ref{sec:TheMapRho}. We can then write\footnote{Disregarding so-called Schwinger terms---here terms proportional to $i[\partial_\mu \phi(x),\phi(x)]$ for a scalar field $\phi$---or, alternatively, reasoning classically.}
\begin{align}\label{E:K1}
	K_1 &=\sum_{n=1}^N \partial^\mu (\Phi_n)_k^\dag G_\mu (\Phi_n)_k + (\Phi_n)_k^\dag G^{\mu \dag} 
	\partial_\mu (\Phi_n)_k \nn \\
	&=\sum_{n=1}^N \rho(\partial^\mu (\Phi_n)_k^\dag) \rho(G_\mu) \rho((\Phi_n)_k) + \rho((\Phi_n)_k^\dag) \rho(G^{\mu \dag}) 
	\rho(\partial_\mu (\Phi_n)_k) \nn \\
	  &= \sum_{n=1}^N \partial^\mu \Phi_n^T \mc{T}_\mu \Phi_n + \Phi_n^T(- \mc{T}^{\mu}) 
	\partial_\mu \Phi_n
\end{align}
  where the subscript $k$ in $(\Phi_n)_k$ indicates this is the usual complex Higgs $k$-plet (in the case $k=2$ the usual complex Higgs doublet), while $\Phi_n$
  is the $2k$ dimensional real vector 
\begin{align}\label{E:PhiReal}
   \Phi_n=\left(\begin{array}{c}\Psi_n \\ \Theta_n
 \end{array}\right),
\end{align}
where $(\Phi_n)_k=\Psi_n+i \Theta_n$, and where we also use
  eqs.~\eqref{Def:rhov}, \eqref{Def:rhovdag} and \eqref{E:rhoreal}.

 In eq.~\eqref{E:K1} we have applied the transformation $\rho$ 
  on the gauge terms $G^\mu$ defined in eq.~\eqref{def:G},
\begin{align}\label{E:rhoT}
	\rho \left( G^\mu \right) &= \mc{T}^\mu,\quad \rho\, ( G^{\mu \dag} ) = - \mc{T}^\mu,
\end{align}
where $\mc{T}$ then reads
\begin{align}\label{E:kinetic1}
	{\cal T^\mu}= \left(\begin{array}{cc}
   gW^\mu_I &-gW^\mu_R-g'\,YB^\mu\\
   gW^\mu_R+g'YB^\mu&gW_I^\mu
 \end{array}\right),
\end{align}
with $W^\mu_R = \sum_{i}' W^\mu_i\, T^{\text s}_i$, summed over the set of
real symmetric generators $T^s_i$ of $SU(k)$, and $W^\mu_I =
\text{i}\sum_{i}' W^\mu_i \,T^{\text a}_i$, summed over the set of imaginary
antisymmetric generators $T^a_i$.  For $k=2$ the two sets are respectively
$\left\{\frac{1}{2}\sigma^1,\frac{1}{2}\sigma^3\right\}$ and
$\left\{\frac{1}{2}\sigma^2\right\}$.

Finally, we consider the kinetic terms quadratic in the gauge fields,
\begin{align}\label{E:K2}
	K_2 &= \sum_{n=1}^N (\Phi_n)_k^\dag G^{\mu \dag}  G_\mu (\Phi_n)_k \nn \\
	&= \sum_{n=1}^N \rho((\Phi_n)_k^\dag)\rho( G^{\mu \dag} )\rho( G_\mu)\rho( (\Phi_n)_k) \nn \\
	&= -\sum_{n=1}^N \Phi_n^T \mc{T}^2 \Phi_n.
\end{align}

\paragraph{The symmetries of $K_0$} 
When we first ignore couplings to the gauge fields the remaining terms can be written
\begin{equation}\label{E:kinetic0}
  K_0 = \sum_{n=1}^N \sum_{\alpha=1}^k 
  \left(\partial^\mu\Psi_{n\alpha}\,\partial_\mu\Psi_{n\alpha}
  +\partial^\mu\Theta_{n\alpha}\,\partial_\mu\Theta_{n\alpha}\right),
\end{equation}
where the group index $\alpha$ labels the $k$ components of
$\Phi_n=\Psi_n+\text{i}\Theta_n$.  This term is invariant under rotation of
{\em all\/} components $\left\{\Psi_{n\alpha}, \Theta_{n\alpha}\right\}$ into
each other. I.e., the symmetry group of $K_0$ is $O(2kN)$. The connected part
of this group is $SO(2kN)$, whose generators are all real antisymmetric
matrices, $L_{mn,\alpha\beta}=-L_{nm,\beta\alpha}$ (i.e.\ $L^T=-L$, where
transposition refers to both sets of indices).  
\paragraph{The symmetries of $K_0$ \emph{and} $K_1$:} Next,
consider the terms linear in the gauge fields again, cf.~eq.~\eqref{E:K1}, 
\begin{equation}\label{E:K12}
 K_1 = \sum_{n=1}^N \left(\left(\partial_\mu \left( 
 \Psi_n^T, \Theta_n^T\right)\right)
 {\cal T}^\mu
 \left(\begin{array}{c}\Psi_n\\\Theta_n
 \end{array}\right)-\left( 
 \Psi_n^T, \Theta_n^T\right)
 {\cal T}^\mu
 \partial_\mu
 \left(\begin{array}{c}\Psi_n\\\Theta_n
 \end{array}\right)\right),
\end{equation}
with ${\cal T}$ given in eq.~\eqref{E:kinetic1}.

 Consider now an infinitesimal
transformation \( \delta \Phi_{m,\alpha} = L_{mn,\alpha\beta}\,
\Phi_{n,\beta}\), and ${\cal T}$ denoting the $2k\times 2k$ antisymmetric
matrix in equation~\eqref{E:kinetic1} (in group indices $\alpha,\,\beta$ ---
in addition it is proportional to the $N\times N$ unit matrix in row indices).
The requirement that this is an infinitesimal symmetry transformation for
$K_1$ is that $L^T\,{\cal T}+{\cal T}\,L=0$.  Or, when we restrict $L$ to be
antisymmetric so that it also is an infinitesimal symmetry transformation for
$K_0$,
\begin{equation}\label{E:L-constraint}
 L_{mn,\alpha\beta}\, {\cal T}_{\beta\gamma} 
- {\cal T}_{\alpha\beta}\,L_{mn,\beta\gamma} = 0.
\end{equation} 
In order to determine the allowed structure of $L$, we expand these matrices
into terms of definite symmetries ($L^{({\text s})}$ symmetric,
and $L^{({\text a})}$ antisymmetric)
in the $mn$ indices:
\begin{equation} \label{E:L-expansion}
L_{mn,\alpha\beta}
= \sum \left( S_{\alpha\beta}L_{mn}^{({\text a})}
+A_{\alpha\beta}L_{mn}^{({\text s})}\right),
\end{equation}
with $S$ (symmetric) and $A$ (antisymmetric) restricted by the constraint
\eqref{E:L-constraint}. The sum runs over all possible combinations of allowed
matrices\footnote{Without the restriction \eqref{E:L-constraint} there would
be $\frac{1}{2}k(2k+1)N(N-1)+\frac{1}{2}k(2k-1)N(N+1)= kN(2kN-1)$ independent
terms, equal to the number of generators of $SO(2kN)$.}.  We next note that
the antisymmetric matrices ${\cal T}$ can be expanded in the set
\begin{equation}\label{E:basis}
 \widehat{\cal T} =   \left\{ 
 \,T^{\text a}_i\,\,{\cal I},\, T^{\text s}_i\,{\cal J},\,{\cal J}\,
 \right\} = 
 \left\{
 \begin{pmatrix}T^{\text a}_i&0_k\\0_k&T^{\text a}_i\end{pmatrix},
 \begin{pmatrix}0_k&T^{\text s}_i\\-T^{\text s}_i&0_k\end{pmatrix},
 \begin{pmatrix}0_k&1_k\\-1_k&0_k\end{pmatrix}
 \right\}.
\end{equation}
By substituting \eqref{E:L-expansion} into \eqref{E:L-constraint} we are led
to search for the set of $2k\times 2k$ real matrices $S$ and $A$ which commute
with ${\cal T}$ for arbitrary values of the fields $W_i^\mu$ and $B^\mu$. It
is sufficient to verify that this property holds for all elements of the set
$\widehat{\cal T}$.  Let
\begin{equation} \label{E:X-expansion}
X=\begin{pmatrix}X_{11}&X_{12}\\X_{21}&X_{22}\end{pmatrix},
\quad
X=S, \quad \text{or} \quad 
X=A.
\end{equation}
Requiring commutativity [see eqs.~(\ref{E:L-constraint}) and
(\ref{E:L-expansion})] with the three types of matrices in $\widehat{\cal T}$
we obtain the conditions
\begin{subequations}
\begin{align}
  &X_{jk} T^{\text a}_i = T^{\text a}_i X_{jk}, \label{E:TA}\\
  &X_{11} T^{\text s}_i = T^{\text s}_i X_{22},\quad
  X_{22} T^{\text s}_i = T^{\text s}_i X_{11},\quad
  X_{12} T^{\text s}_i = -T^{\text s}_i X_{21},\quad
  X_{21} T^{\text s}_i = -T^{\text s}_i X_{12},\label{E:TS}\\
  &X_{11} = X_{22},\quad
  X_{12} = -X_{21}.\label{E:TJ}
\end{align}
\end{subequations}
Using \eqref{E:TJ} we find that $X_{11}$ and $X_{12}$
must commute with all matrices $T_i^{\text a}$, $T_i^{\text s}$
(assumed to form an irreducible representation). By Schur's
lemma they must then be proportional to the $k\times k$ unit matrix,
so that $S\propto {\cal I}$ and $A\propto{\cal J}$. 
Thus, the Lie algebra of the symmetry
group of $K_0$ and $K_1$ consists of elements of the form
\begin{equation}\label{E:LieGenerators}
  L_{mn} = {\cal I}\, L_{mn}^{(\text a)} + {\cal J}\,L_{mn}^{(\text s)}.
\end{equation}
This is the Lie algebra of $U(N)$ written in real variables.

\paragraph{The symmetries of $K_0$ \emph{and} $K_1$ in the limit $g' \to 0$:} 
A more interesting situation arises if we remove ${\cal J}$ from the set
$\widehat{\cal T}$, as would apply to the limit
$g'\to0$.  Then we still find that $X_{11}+X_{22}$ and $X_{12}-X_{21}$ must
commute with all matrices $T_i^{\text a}$, $T_i^{\text s}$, and hence must be
proportional to the unit matrix. Further, $X_{11}-X_{22}$ and $X_{12}+X_{21}$
must commute with all $T_i^{\text a}$, but anticommute with all $T_i^{\text
s}$. For $k=2$, i.e.\ for the gauge group $SU(2)_L\times U(1)_Y$ in the limit
$g'\to 0$, we find that nonzero solutions of (\ref{E:L-constraint}),
\begin{equation}\label{eq:matrix_varepsilon}
  X_{11}-X_{22}\propto \varepsilon \equiv \text{i}\,\sigma^2,\quad
  X_{12}+X_{21}\propto \varepsilon,
\end{equation}
are possible [see eqs.~(\ref{E:L-expansion}) and (\ref{E:X-expansion})]. This
means that the possible antisymmetric matrices $A$ may be any linear
combination of matrices from the set
\begin{equation}
  {\cal G} = \left\{\,\begin{pmatrix}0&\varepsilon\\
    \varepsilon&0\end{pmatrix}\,,\,
  \begin{pmatrix}\varepsilon&0\\0&-\varepsilon\end{pmatrix}\,,\,
  {\cal J}\,\right\},
\end{equation}
where the $2 \times 2$ matrix $\varepsilon$ was defined in
eq.~\eqref{eq:matrix_varepsilon}.  The set ${\cal G}$ is a basis of generators
for $SU(2)$. Thus, the Lie algebra of symmetry generators for $K_0$ and $K_1$
in this case consists of elements of the form
\begin{equation}\label{eq:RealLieAlgSpN}
  L_{mn} = {\cal I}\, L_{mn}^{(\text a)} 
+ \sum_{A\in{\cal G}} A\,L_{mn}^{(\text s)},
\end{equation}
allowing all possible symmetric $N\times N$ matrices $L^{(\text s)}$ for each
$A$.  There are $\frac{1}{2}N(N-1)+\frac{3}{2}N(N+1)=2N^2+N$ independent
terms, equal to the number of generators of the $N\times N$ quaternionic
symplectic group $Sp(N)$. The generators~\eqref{eq:RealLieAlgSpN} generate
$Sp(N)$, where the elements of ${\cal G}$ act as the
quaternions $i$, $j$ and $k$.

The above results were again found under the assumption that the fields $W^\mu_i$
are arbitrary, and kept constant under the transformation.  Combined
$SU(k)$ transformations of the $W^\mu_i$ and the $\Phi_m$ fields still remain a
symmetry.
 This symmetry is enlarged to
$SU(k)\times Sp(N) $ as $g'\to 0$. (In the case $g'\ne 0$ it is $SU(k)\times U(N)$.)
In the case $k=2$ and $g'\to 0$ the custodial $SO(4)$ symmetry\footnote{The custodial $SO(4)$ symmetry cannot be extended to an $O(4)$ symmetry \cite{PhD}.} is contained in $SU(2)\times Sp(N)$ in the following way: $SO(4)\cong SU(2)_L\times SU(2)_R \subseteq SU(2)_L \times Sp(N)$.\footnote{We are grateful to H.~Haber, J.~P.~Silva and P.~Ferreira for pointing out a mistake at this point in a previous manuscript.} (More presicely, $SO(4)\cong (SU(2)_L\times SU(2)_R)/\N{Z}_2$ \cite{Cornwell:1985xt}.)
The group $SU(2)_R\subseteq Sp(N)$ is generated (through exponentiation) by choosing
$L^{(\text a)}=0_{N\times N}$ and each $L^{(\text s)}\propto I_{N\times N}$ in eq.~\eqref{eq:RealLieAlgSpN}. The generators of $SO(4)$ are hence the 3 generators in ${\cal G}$ plus the 3 generators of the $SU(2)_L$ gauge group
(written in real form). Finally, the $U(1)_Y$ hypercharge symmetry group is contained in $SU(2)_R$ \cite{Willenbrock:2004hu}.
 
 In the more general case of an $SU(k)\times Sp(N)$ symmetry of $K_0$ and $K_1$ in the limit $g'\to 0$ there is, in the same manner as above, a ``custodial'' $SU(k)\times SU(2)_R$ symmetry, which also will contain $SU(2)\times SU(2)_R \cong SO(4)$ subgroups.

\paragraph{The symmetries of $K_2$}
Next, consider the terms quadratic in the gauge fields cf.~eq.~\eqref{E:K2},
\begin{equation}\label{E:K22}
 K_2 = -\sum_{n=1}^N \left( 
 \Psi_n^T, \Theta_n^T\right)
 {\cal T}^2
 \left(\begin{array}{c}\Psi_n\\\Theta_n
 \end{array}\right).
\end{equation}
As in the symmetry analysis of
$K_1$ we want to find all matrices $X$ such that $X {\cal T}^2 = {\cal T}^2
X$. All matrices $X$ which commute with $\widehat{\cal T}$ will fulfill this
criterion (since ${\cal T}^2$ can be expanded in a set which consists of
products of all possible {\it pairs} of matrices from $\widehat{\cal
T}$). Therefore, the symmetries of $K_1$ are also symmetries of $K_2$.

\section{Spontaneous symmetry breakdown}
\label{sec:ssb}
In this section we return to the case of $k=2$, i.e.\ with $SU(2)_L\times
U(1)_Y$ as the gauge group and a row of $N$ Higgs doublets $\Phi_m$. Note
however that many of our considerations have straightforward generalizations
to $k>2$.

As for the Standard Model, the potential $V_\text{g}$ of equation
\eqref{E:potNHDM}, or $V_C$ of equation \eqref{E:potNHDMCPinv}, may
acquire its minimum at nonzero values of the scalar fields, $\langle \Phi
\rangle_0 = \Phi^{(0)}$, where $\Phi$ (without a lower index) refers to the
whole set of fields $\Phi_m$.  This point, $\Phi^{(0)}$, will belong to one or
more manifolds of equivalent minima related by the symmetries of the
potential. One may use these symmetries to transform $\Phi^{(0)}$ to a
particular form. A possible one is to require for $\Phi^{(0)}_1$ that only its
lowest real component is nonzero.  This can always be achieved by an
$SU(2)_L\times U(1)_Y$ gauge transformation.  Next, the upper component of
$\Phi^{(0)}_2$ can be made real by the remaining $U(1)$ gauge transformation
which keeps $\Phi^{(0)}_1$ unchanged. Then one has no gauge freedom left to
change $\Phi^{(0)}_n$ for $n\ge3$. However, it was shown by Barroso et
al.~\cite{Barroso:2006pa} 
that a sequence of unitary row transformations can
shift the vacuum expectation values to the first two fields of the row
only\footnote{One may collect the quantities $\Phi^{(0)}_{m\alpha}$
($\alpha=1,2$) into two $N$-component complex vectors $\tilde{\Phi}^{(1)}$ and
$\tilde{\Phi}^{(2)}$. By a $U(N)$ row transformation one may first rotate
$\tilde{\Phi}^{(1)}_m$ so that only the component $\tilde{\Phi}^{(1)}_1$ is
nonzero, with $\tilde{\Phi}^{(1)}_1$ real.  There is a group of $U(N-1)$
transformations preserving this condition; this may be used to transform
$\tilde{\Phi}^{(2)}_m$ so that only the components $\tilde{\Phi}^{(2)}_1$,
$\tilde{\Phi}^{(2)}_2$ are nonzero, with $\tilde{\Phi}^{(2)}_2$ real. One
cannot do better due to the existence of four real $U(N)$ invariant parameters
in $\vert|\tilde{\Phi}^{(1)}\vert|$, $\vert|\tilde{\Phi}^{(2)}\vert|$, and
$\tilde{\Phi}^{(1)\dag}\, \tilde{\Phi}^{(2)}$. But there remains a $U(N-2)$
group of transformations preserving this condition which can be used for other
purposes. For an $SU(k)\times U(1)$ gauge group one may generalize this
procedure to $k$ vectors $\tilde\Phi^{(j)},\;j=1\ldots k$.}, for instance
(when written in complex form)
\begin{equation}\label{E:HiggsbasisVEV}
 \Phi^{(0)}_1 = \left(\begin{array}{c}0\\
   v_1\end{array}\right),\quad
 \Phi^{(0)}_2 = \left(\begin{array}{c}u_2\\
   v_2\text{e}^{\text{i}\delta}\end{array}\right),\quad
 \Phi^{(0)}_n = 0 \text{ for $n\ge3$,}
\end{equation}
with $v_1$, $u_2$, $v_2$ and $\delta$ real. The special case $u_2=\delta=0$ is
usually referred to as {\em vacuum alignment\/}, in which case we may also
transform $v_2$ to zero by an orthogonal row transformation involving only
$\Phi_1$ and $\Phi_2$.  This is known as the {\em Higgs basis\/}
\cite{HiggsBasis}.  However, for other purposes it may be more convenient to
adopt a ``democratic'' basis in which the lower component of all (or most)
fields $\Phi_m$ have a nonzero real expectation value. It is related to the
Higgs basis by an orthogonal row transformation which preserves the form of
$C$ and $U(1)$ electromagnetic gauge transformations (the latter
preserving the definition of electric charge).

Assume now the case of vacuum alignment and a potential $V_{O(4)}$ which
is $O(4)$ invariant. Then the existence of the vacuum expectation values
$\Phi^{(0)}$ will break the (explicit) symmetry down to $O(3)$, with the
consequence that the Higgs boson particle spectra and other physical
properties will organize themselves into multiplets of $O(3)$ (broken by
perturbative corrections in $g'$). The number of broken symmetry generators is
3 whether we consider the symmetry broken from $U(2)$ to $U(1)$ or from $O(4)$
to $O(3)$; this leads to the existence of 3 Higgs ghosts and no extra (pseudo-)
Goldstone bosons\footnote{This remains true for general values of $k\ge2$; a
set of aligned vacuum expectation values will break $U(k)$ to $U(k-1)$ and
$O(2k)$ to $O(2k-1)$. The number of broken generators is $2k-1$ in both
cases. The situation is different if we have {\em two\/} broken real
directions, as in equation \eqref{E:HiggsbasisVEV} with $u_2=0$ but $\delta
\ne 0$.  Cf. section~\ref{sec:NonAligned}.}.

\subsection{Mass-squared matrices}
\label{sec:MassSquaredMatrices2}

To make these statements slightly more explicit, as needed for calculation of
the zero'th order (in $g$ and $g'$) particle masses, we expand the potential
around $\Phi^{(0)}$ to second order.
There are no first order terms since we are expanding
around a minimum.  The matrix of second derivatives is the mass-squared matrix
$M^2_{mn\alpha\beta}$.  It is restricted by symmetries to have a block
diagonal form in the group indices $\alpha$, $\beta$. We use coordinates where
$\Phi_m=\Psi_m+\text{i}\Theta_m$ is expressed in terms of four real fields,
\begin{equation}\label{Eq:2-comp-notation}
 \Phi_m = \Psi_m + \text{i}\Theta_m =
 \left(\begin{array}{cc}\phi_{m1}+\text{i}\phi_{m2}\\
   v_m+\eta_m+\text{i}\phi_{m3}
   \end{array}\right),
 \quad \Phi^{(0)}_m = \left(\begin{array}{c}0\\v_m\end{array}\right).
\end{equation}
It is now convenient to represent these on real form as
\begin{equation} \label{Eq:4-comp-notation}
\Phi_m = 
\begin{pmatrix}
\Psi_{m} \\ \Theta_m
\end{pmatrix}
=
\begin{pmatrix}
\phi_{m1} \\ v_m+\eta_m \\ \phi_{m2} \\ \phi_{m3}
\end{pmatrix}.
\end{equation}
We have the expansion
\begin{equation}\label{E:TaylorExpansion}
   V(\Phi^{(0)}+\Delta\Phi) = \langle V\rangle _0 +
   \frac{1}{2} \left\langle\frac{\partial^2 V}{\partial \Phi_{m\rho}\,
     \partial\Phi_{n\sigma}}\right\rangle_0\,
   \Delta\Phi_{m\rho}\Delta\Phi_{n\sigma} + {\cal O}(\Delta\Phi^3),
\end{equation}
where $\Phi_{m\rho}$ denotes one of the {\em four\/} possibilities
$\phi_{m1}$, $\eta_{m}$, $\phi_{m2}$, $\phi_{m3}$, and the subscript $0$
indicates that a quantity is evaluated at $\Phi=\Phi^{(0)}$.  Now a set of
generators for $SO(4)$\footnote{Equivalently $O(4)$: $SO(4)$ and $O(4)$ have
the same Lie algebra and hence the same generators.} is
\begin{align}
 &J_1 = \left(\begin{array}{rrrr}
   0&1&0&0\\-1&0&0&0\\0&0&0&0\\0&0&0&0
 \end{array}\right),\quad
 J_2 = \left(\begin{array}{rrrr}
   0&0&0&0\\0&0&-1&0\\0&1&0&0\\0&0&0&0
 \end{array}\right),\quad
 J_3 = \left(\begin{array}{rrrr}
   0&0&0&0\\0&0&0&-1\\0&0&0&0\\0&1&0&0
 \end{array}\right),\nonumber\\
 &J_4 =\left(\begin{array}{rrrr}
   0&0&0&0\\0&0&0&0\\0&0&0&1\\0&0&-1&0
 \end{array}\right), \quad
 J_5 = \left(\begin{array}{rrrr}
   0&0&0&-1\\0&0&0&0\\0&0&0&0\\1&0&0&0
 \end{array}\right),\quad
 J_6 =\left(\begin{array}{rrrr}
   0&0&-1&0\\0&0&0&0\\1&0&0&0\\0&0&0&0
 \end{array}\right),
\end{align}
where $J_1,J_2,J_3$ will transform the vacuum expectation value $\Phi^{(0)}$,
while $J_4,J_5,J_6$ leave it unchanged, cf.~Eq.~(\ref{Eq:4-comp-notation}).
In terms of these, the broken generators of the $SU(2)\times U(1)_Y$ gauge
group, written in real form by the transformation 
 $\rho$ defined in~\eqref{Def:rho}, are 
\begin{equation}\label{E:SU2gens}
 \frac{\text{i}}{2}\sigma^1 \to \frac{1}{2} (J_2+J_5),\quad 
 \frac{\text{i}}{2}\sigma^2 \to \frac{1}{2} (J_1+J_4),\quad
 \frac{\text{i}}{2}(1-\sigma^3) \to J_3, 
\end{equation}
and the unbroken $U(1)$ (electromagnetic gauge) generator is
\begin{equation}
 \frac{\text{i}}{2}(1+\sigma^3) \to J_6.
\end{equation}
If now $V$ is invariant under
\begin{equation*}
 \Delta\phi_{m1}\to -\Delta\phi_{m1},\quad 
  \Delta\eta_{m} \to \Delta\eta_{m}, \quad
 \Delta\phi_{m2} \to -\Delta\phi_{m2},\quad
 \Delta\phi_{m3}\to \Delta\phi_{m3}, 
\end{equation*}
there can be no terms in \eqref{E:TaylorExpansion} mixing the sets
$\left\{\Delta\phi_1,\Delta\phi_2\right\}$ and
$\left\{\Delta\eta_m, \Delta\phi_{m3}\right\}$. If $V$ in addition is invariant
under
\begin{equation*}
 \Delta\phi_{m1}\to \Delta\phi_{m2},\quad 
 \Delta\eta_{m} \to \Delta\eta_{m},\quad 
 \Delta\phi_{m2} \to -\Delta\phi_{m1},\quad
 \Delta\phi_{m3}\to \Delta\phi_{m3},
\end{equation*}
there can be no terms in \eqref{E:TaylorExpansion} mixing $\Delta\phi_{m1}$
and $\Delta\phi_{m2}$, and we must have
\begin{equation}\label{def:Mch}
 \left\langle\frac{\partial^2 V}{\partial \phi_{m1}\,\partial\phi_{n1}}
\right\rangle_0 
 =\left\langle\frac{\partial^2 V}{\partial \phi_{m2}\,\partial\phi_{n2}}
\right\rangle_0 
 \equiv M^{2}_{\text{ch},mn}.
\end{equation}
We refer to this as the {\em charged\/} mass-squared matrix.  The
transformations considered generate a $\N{Z}_4$ subgroup of the $U(1)$ gauge group
generated by $J_6$, assumed to be a symmetry of $V$. We have formulated it
this way as a reminder that invariance under discrete subgroups may be
sufficient to impose useful restrictions on the mass matrices. 

If $V$ is invariant under $C$ transformations,
\begin{equation}
 \Delta\eta_m \to \Delta\eta_m,\quad 
 \Delta\phi_{m3}\to -\Delta\phi_{m3},
\end{equation}
(irrespective of how we define $C$ to operate on the charged sector,
e.g.~$\Delta\phi_{m1}\to \Delta\phi_{m1}$, $\Delta\phi_{m2}\to -\Delta\phi_{m2}$)
there can be no terms in \eqref{E:TaylorExpansion} mixing $\Delta\eta_{m}$
and $\Delta\phi_{m3}$.  Thus the neutral mass-squared matrix decomposes into
two more blocks, a {\em$C$ even\/} and a {\em $C$ odd\/}
one,
\begin{equation}
 M^2_{C+,mn}= \left\langle\frac{\partial^2 V}{\partial \eta_{m}\,
   \partial\eta_{n}}\right\rangle_0,\quad
 M^2_{C-,mn}= \left\langle\frac{\partial^2 V}{\partial \phi_{m3}\,
   \partial\phi_{n3}}\right\rangle_0.
\end{equation}
If $V$ in addition is invariant under the transformations
\begin{equation}\label{eq.<-lambda3=0}
 \Delta\phi_{m1} \to \Delta\phi_{m1},\quad 
 \Delta\eta_{m} \to \Delta\eta_{m},\quad
 \Delta\phi_{m2} \to -\Delta\phi_{m3},\quad 
 \Delta\phi_{m3} \to \Delta\phi_{m2},
\end{equation} 
which generate a $\N{Z}_4$ subgroup of the $SO(2)$ symmetry group
generated by $J_4$,
we obtain the relation
\begin{equation}\label{E:massDeg}
   M^2_{C-,mn}= M^2_{\text{ch},mn}.
\end{equation}
This explicitly demonstrates mass degeneracy between the charged and the
$C$ odd sectors~\cite{Solberg}. Especially, if the potential is $O(4)$-invariant \eqref{E:potNHDMO2kinv}, that is, we have a $C$-invariant theory where also
the parameters $\lambda_{ab}^{(3)}=0$ [the latter implies
\eqref{eq.<-lambda3=0}]\footnote{For supersymmetric theories we
typically have $\lambda_{ab}^{(3)}\ne 0$.}, the above symmetry criteria for mass degeneracy are
valid. Moreover, since the renormalization
is not changed when the Higgs fields acquire a vacuum expectation value \cite{Lee:1968da}, we do not
get any mass renormalization counterterms from the quartic operators. So even though $O(4)$-violating quartic terms
proportional to $\lambda_{ab}^{(3)}$ cannot alone be prohibited by any discrete symmetry imposed on the NHDM Lagrangian \cite{PhD}, they will not show up as counterterms when renormalizing the masses. Hence, the mass degeneration \eqref{E:massDeg} will only be broken by loop corrections involving gauge bosons, since we get an exact $SO(3)$ symmetry
when $g'= 0$. With $g'\ne 0$ and
hence with an approximate
$SO(3)$ symmetry, the mass differences of the charged and $CP$-odd sectors will be of order $\mc{O}(g'^4)\propto \mc{O}(e^4)$. 

 On the other hand, the $SO(3)$ symmetry 
between $CP$-odd and charged sectors 
could also be broken 
by counterterms of the type $\lambda_{ab}^{(3)} \widehat{C}^2$, even though these terms are set to zero in the original potential, if we are
considering scattering processes and not mass relations.

\subsection{The Higgs ghosts}
\label{sec:VEVsAlignedInTheEtaDirection}

Let $\Delta\Phi$ be chosen so that $\Phi^{(0)}+\epsilon\Delta\Phi+{\cal
O}(\epsilon^2)$ is a family of minima related by the symmetry of the potential
$V$,
\begin{equation}
 \frac{\partial}{\partial\Phi_{m\alpha}} V(\Phi^{(0)}+\epsilon\Delta\Phi)=0,
\end{equation}
to first order in $\epsilon$.  By differentiating this relation with respect
to $\epsilon$ and then setting $\epsilon=0$ we find
\begin{equation}
 \left\langle\frac{\partial^2 V}{\partial\Phi_{m\alpha}\,
   \partial\Phi_{n\beta}}\right\rangle_0 \Delta\Phi_{n\beta} = 0,
\end{equation}
which reflects the fact that the matrix $M^2_{mn\alpha\beta}$ has zero
eigenvalues with corresponding eigenvectors $\Delta\Phi_{n\beta}$. We may
take the latter to be $\Delta\Phi^{(i)} \propto J_i \Phi^{(0)}$ for
$i=1,2,3$. Normalized,
\begin{equation}
 \Delta\Phi^{(1)}_m = \left(v_m,0,0,0\right)^T/a,\quad
 \Delta\Phi^{(2)}_m = \left(0,0,v_m,0\right)^T/a,\quad
 \Delta\Phi^{(3)}_m = \left(0,0,0,v_m\right)^T/a,
\end{equation}
with $a^2 = \sum_m v_m^2$.  The massless excitations in these directions
correspond to a triplet of Higgs ghosts. There will be $N-1$ additional
$SO(3)$ triplets of excitations in directions orthogonal to the ghosts.  They
correspond to physical particles. There will also be $N$ $SO(3)$ singlets,
transforming evenly under $C$, corresponding to physical particles.  In
the case of $N=2$, the triplet is ($H^+$, $H^-$, $A$), whereas the singlets
are $h$ and $H$ \cite{Gunion:1989we}.

\subsection{Non-aligned vacuum expectation values}
\label{sec:NonAligned}

We have assumed vacuum alignment in much of the previous discussions of this
section.  The phenomenologically most realistic deviation from this case is
that we have a situation with {\em two\/} (real) broken directions, as in
\eqref{E:HiggsbasisVEV} with $u_2=0$ but $\delta \ne 0$.  This corresponds to
a situation which preserves the $U(1)$ electromagnetic gauge symmetry, generated by $J_6$.
Its corresponding definition of electric charge is preserved, but the $C$ symmetry
(or, here equivalently\footnote{For the Lagrangian \eqref{E:fullNHDMpot}, spontaneous $C$ and $CP$ violation (SCV and SCPV) are equivalent: By definition \cite{Branco:1999fs}, $CP$ ($C$) is broken spontaneously if (1) There is a transformation that can be physically interpreted as $CP$ ($C$) and which keeps the Lagrangian invariant and (2) There is no transformation that can be physically interpreted as $CP$ ($C$) which keeps both the Lagrangian and the vacuum invariant. \newline
{\bf (SCV $\Rightarrow$ SCPV):} Assume $C$ is spontaneously broken, and implement $P$ by the spatial reflection 
\begin{align}\label{E:trivP}
	{\cal P} \Phi_n(t,\boldsymbol{r}) {\cal P}^\dag = \Phi_n(t,-\boldsymbol{r}).
\end{align}
Hence condition (1) of SCPV is satisfied. Next, assume condition (2) for SCPV is not satisfied. Then there is a $CP$ transformation
\begin{align}\label{E:nonSCPV}
	({\cal C P}) \Phi_n(t,\boldsymbol{r}) ({\cal C P})^\dag = U_{m n}^{CP}\Phi_n^{\dag \, T} (t,-\boldsymbol{r}),
\end{align}
  which leaves both the Lagrangian and the vacuum unaltered. But then
  the $C$ transformation given by
\begin{align}
	{\cal C} \Phi_n(t,\boldsymbol{r}) {\cal C}^\dag = U_{m n}^{C}\Phi_n^{\dag \, T} (t,\boldsymbol{r}), \nn
\end{align}
  with $U_{m n}^{C}=U_{m n}^{CP}$ will infer that condition (2) of SCV does not hold, since
  the spatial reflection \eqref{E:trivP} does not change the vacuum nor the physics of the Lagrangian. This is a contradiction, and hence
  also condition (2) of SCPV must hold.
    \newline
{\bf (SCPV $\Rightarrow$ SCV):} Conversely, if $CP$ is spontaneously broken and condition (1) hence is satisfied by  
\[
({\cal C P}) \Phi_n(t,\boldsymbol{r}) ({\cal C P})^\dag = U_{m n}^{CP}\Phi_n^{\dag \, T} (t,-\boldsymbol{r}),
\]
 we can re-implement the matrix $U^C$ as $U^{CP}$ above, and re-implement $P$ as the trivial transformation \eqref{E:trivP}, and hence condition (1) of SCV is satisfied. Assume condition (2) of SCV does not hold. Then we, in a similar manner as for the case (SCV$\Rightarrow$SCPV), can let $U^{CP}=U^C$ and define $P$ as in \eqref{E:trivP}, and hence SCPV does not hold either, which is a contradiction. Hence condition (2) of SCV is satisfied.}, $CP$ symmetry) is spontaneously broken. In this situation, assuming the potential only has the $SU(2)\times U(1)$ gauge symmetry, the three $SU(2)$ generators are spontaneously broken, and we hence only get three Higgs ghosts and no (pseudo-) Goldstone bosons. 

When we only have an $SU(2)\times U(1)$ symmetry in the potential, operators of the type $\widehat{C}$
are present 
 in the quadratic or in the quartic part of the potential. Two of the excitations of the charged mass squared matrix are still massless, corresponding to the usual charged
Higgs ghosts.\footnote{The relation between the complex and real formulations of the charged mass squared matrix is given by the map $\rho$ of appendix \ref{sec:TheMapRho}:
Let $M^2_{\text{c}}$ denote the $N\times N$ complex mass squared matrix, and let $M^2_{\text{r}}$ 
be the corresponding $2N\times 2N$ real matrix. They are related by 
$$\phi^{-\text{T}} M^2_{\text{c}}\phi^+ = (\phi_1\,\phi_2)\rho(M^2_{\text{c}})\begin{pmatrix}\phi_1\\ \phi_2\end{pmatrix},$$ 
as can be seen from eq.~\eqref{E:rhoreal}. Then
 $\rho(M^2_{\text{c}})=M^2_{\text{r}}$. The matrix $M^2_{\text{c}}$ is Hermitian, and hence has only real eigenvalues. It then follows from the definition of $\rho$ that $\lambda$ is an eigenvalue of $M^2_{\text{c}}$ if and only if $\lambda$ is an eigenvalue of $M^2_{\text{r}}$. Moreover, if $v$ is an eigenvector of $M^2_{\text{c}}$, $(\Re v,\Im v)^\text{T}$ and $(\Im v, -\Re v)^\text{T}$ will be eigenvectors of $M^2_{\text{r}}$ with the same eigenvalue $\lambda$. Hence oppositely charged particles will have the same mass, although
 terms $\phi_{1m}$ and $\phi_{2n}$ mix by the presence of operators $\widehat{C}_{mn}$ in
 the quadratic part of the potential or by complex (i.e.~non-aligned) VEVs, and hence violate $C$. (This mass degeneration is a consequence of the remaining $U(1)\cong SO(2)$ hypercharge symmetry, i.e.~that the generator $J_6$ is unbroken.) The identification \eqref{def:Mch} assumed that $\phi_{1m}$ and $\phi_{2n}$ did not mix, i.e.~that $\text{Im}(M^2_{\text{c}})=0$, cf.~\eqref{Def:rho}.}

In the case of non-aligned VEVs (or operators $\widehat{C}$ present in the quadratic part of the potential) the $C$ even and odd
excitations generally mix to give a $2N \times 2N$ mass squared matrix for the neutral
particles. Here one of the excitations will be massless, corresponding to a neutral
Higgs ghost. 

On the other hand, assume that $V$ is invariant under $O(4)$
transformations. Then there are no operators of the type $\widehat{C}$ present in the potential. The explicit $O(4)$ symmetry is now broken down to $O(2)\simeq
U(1)$, so that 5 generators are broken. As before, 3 of these will generate
excitations which correspond to the Higgs ghosts; the remaining 2 will
correspond to nearly massless charged
pseudo-Goldstone\footnote{Pseudo-Goldstone bosons stem from broken generators
of the extra $O(4)$ symmetry of the potential, while Higgs ghosts by
definition are generated by the broken generators of the gauge symmetry (which
of course is a symmetry of the whole Lagrangian). The pseudo-Goldstone bosons
acquire small masses from radiative corrections, and are hence not massless to
all orders of perturbation theory, like true Goldstone bosons. True Goldstone
bosons are, in contrast, generated by the spontaneous breaking of a symmetry
of a total Lagrangian, not only a potential.} bosons (massless to zero'th
order in $g'$).

To analyze the situation we again write $\Phi = \Phi^{(0)}+\Phi'$ in terms of
real fields,
\begin{equation*}
\Phi_m = \Phi_m^{(0)} + \left(\phi_{m1},\eta_m, \phi_{m2},\chi_{m}\right)^T
\text{ with } \Phi_m^{(0)} = \left(0,v_m,0,w_m\right)^T.
\end{equation*}
$J_4$ and $J_5$ are now also broken by the vacuum expectation values.
Acting with the broken generators on $\Phi^{(0)}$ one finds
five eigenvectors of the mass
matrix with zero eigenvalues, $\Delta\Phi^{(i)} \propto J_i \Phi^{(0)}$. After
normalization
\begin{align}\label{E:normEigenv0}
   &\Delta\Phi^{(1)}_m = \left(v_m,0,0,0\right)^T/a,\nonumber\\
   &\Delta\Phi^{(2)}_m = \left(0,0,-v_m,0\right)^T/a,\nonumber\\
   &\Delta\Phi^{(3)}_m = \left(0,w_m,0,-v_m\right)^T/\sqrt{a^2+b^2},\\
   &\Delta\Phi^{(4)}_m = \left(0,0,w_m,0\right)^T/b,\nonumber\\
   &\Delta\Phi^{(5)}_m = \left(w_m,0,0,0\right)^T/b,\nonumber
\end{align}
where $a^2=\sum_m v_m^2$ and $b^2 = \sum_m w_m^2$. We see that all eigenvectors except $\Delta\Phi^{(3)}_m$ are in the charged sector, hence the two pseudo-Goldstone bosons (i.e.~light Higgs bosons) mentioned above will be charged, as claimed. The eigenvectors in \eqref{E:normEigenv0} are
normalized, but they are not necessarily orthogonal to each other.
Their nonvanishing inner products are
\begin{equation*}
 \left(\Delta\Phi^{(1)},\Delta\Phi^{(5)} \right) 
 = -\left(\Delta\Phi^{(2)},\Delta\Phi^{(4)} \right) 
 = \frac{1}{ab}\sum_m v_m w_m \equiv \cos\vartheta.
\end{equation*}
Here $\vert \sin\vartheta \vert > 0$, since the vacuum expectation values by
assumption are non-aligned.  Thus, the orthonormalized eigenvectors
corresponding to the Higgs ghosts can be written
\begin{align}
 &H^{(1)}_m 
 =\frac{1}{\sqrt{a^2+b^2}}\,\left(v_m,0,w_m,0\right)^T 
 =\frac{a}{\sqrt{a^2+b^2}}\,\Delta\Phi^{(1)}_m 
 + \frac{b}{\sqrt{a^2+b^2}}\,\Delta\Phi^{(4)}_m,\nonumber\\
 &H^{(2)}_m 
 =  \frac{1}{\sqrt{a^2+b^2}}\,\left(w_m,0,-v_m,0\right)^T 
 =\frac{a}{\sqrt{a^2+b^2}}\,\Delta\Phi^{(2)}_m 
 + \frac{b}{\sqrt{a^2+b^2}}\,\Delta\Phi^{(5)}_m,\\
 &H^{(3)}_m 
 =  \frac{1}{\sqrt{a^2+b^2}}\,\left(0,w_m,0,-v_m\right)^T 
 = \Delta\Phi^{(3)}_m,\nonumber
\end{align}
where $H^{(i)}_m \propto G_i \Phi^{(0)}$, $G_i$ denoting the $SU(2)$
generators as given by the map~\eqref{E:SU2gens}.  The two eigenvectors
corresponding to the Goldstone modes are orthogonal to those above,
\begin{align}
 &G^{(1)} =  \frac{1}{\sin\vartheta\,\sqrt{a^2+b^2}}\,
 \left[a\left( \Delta\Phi^{(4)} +\cos\vartheta\, \Delta\Phi^{(2)}\right)
   -b\left( \Delta\Phi^{(1)} 
   -\cos\vartheta\, \Delta\Phi^{(5)} \right)\right],\nonumber\\
 &G^{(2)} =  \frac{1}{\sin\vartheta\,\sqrt{a^2+b^2}}\,\left[
   -a\left(\Delta\Phi^{(5)} - \cos\vartheta\,\Delta\Phi^{(1)}\right) 
   +b\left(\Delta\Phi^{(2)} + \cos\vartheta\,\Delta\Phi^{(4)}\right)
   \right].
\end{align}
They have been orthonormalized. We note that the normalization constant
becomes infinite in the limit of aligned vacuum expectation values,
$\sin\vartheta\to0$. We recall that the set
$\left\{H^{(1)},H^{(2)},H^{(3)},G^{(1)},G^{(2)}\right\}$ are just numerical
eigenvectors of the mass-squared matrix. The corresponding zero mode fields
are the quantum fields obtained by projecting $\Phi'$ on these eigenvectors,
\begin{equation}
  \Phi^{H^{(i)}}_{m\alpha} 
  = \left(H^{(i)},\Phi'\right)\,H^{(i)}_{m\alpha},\quad
  \Phi^{G^{(j)}}_{m\alpha} 
  = \left(G^{(j)},\Phi'\right)\,G^{(j)}_{m\alpha}\quad
  \text{ for $i=1,2,3$ and $j=1,2$.}
\end{equation}
The field $\Phi^{H^{(3)}}$ is the neutral Higgs ghost field, while the fields
$\Phi^{H^{(1)}}$ and $\Phi^{H^{(2)}}$ form the charged Higgs ghost field, and
the fields $G^{(1)}$ and $G^{(2)}$ together form charged Goldstone boson
fields.

If the vacuum expectation values broke the symmetry in even more directions,
as in \eqref{E:HiggsbasisVEV} with both $u_2\ne 0$ and $\delta \ne 0$, the
situation would be different: All 6 generators of $SO(4)$ would be broken,
4 of them corresponding to the 4 broken generators of the $U(2)$ gauge group.
Thus, there would be 2 pseudo-Goldstone bosons also in this case.
\section{Concluding remarks}
\label{sec:conclusion}

We have analyzed the additional (approximate) symmetries
which may arise in multi-Higgs-doublet models, due to the fact that the scalar
potential may have more symmetries than required
by the imposed gauge invariance. Moreover, for the kinetic terms we found that
the symmetry group is $SU(k)\times U(N)$. In the limit $g'\to 0$ 
the symmetry group of the kinetic terms is
enhanced to $SU(k)\times Sp(N)$, which has an $SU(k)\times SU(2)$ subgroup. In the case $k=2$ the latter is the $SU(2)_L\times SU(2)_R \cong SO(4)$ custodial symmetry.
The most general $C$ invariant Higgs 
potential~\eqref{E:potNHDMCPinv} has the same $SO(4)$ symmetry, only broken by the presence of the operator
$\widehat{C}^2$, that is, terms proportional to $\lambda_{ab}^{(3)}$. In the case where $\lambda_{ab}^{(3)}$
is set to zero, we have an exact mass degeneration \eqref{E:massDeg} (assuming vacuum alignment) between 
charged and $C$ odd sectors in the limit $g'\to 0$. When there is no vacuum alignment, 
but rather two broken (real) directions with the electromagnetic generator left unbroken, a pair of light, 
charged Higgs bosons emerge (cf.~section \ref{sec:NonAligned}).  

The introduction of Yukawa couplings could further constrain the theory. With $N$ doublets, one could imagine ``simplified'' models analogous to Model~I and Model~II for the 2HDM, where only one doublet couples to {\it all} fermions, or where some doublets couple to up-type quarks, with others coupling to down-type fermions. Furthermore, with three or more doublets, one could arrange to let each fermion generation couple to its own doublet.

If $n$ doublets are without {\it any} Yukawa couplings, for example due to a discrete $\N{Z}_2$ symmetry, 
\begin{equation}
\Phi_i\to-\Phi_i, \quad i=1,\ldots,n
\end{equation}
then such a sector would provide a dark matter candidate \cite{Barbieri:2006dq}. Indeed, with $n>1$, there would be a whole ``family'' of states in this ``inert'' sector, some of which would carry electric charge.
Those would therefore be observable.

\appendix
\section{$P(k,\N{R})$, the symmetry group of $\widehat{C}^2$}
\label{sec:TheSymmetryGroupOfWidehatC2}
We will here show that the set 
\begin{align}\label{Def:Pk2}
	P(k,\N{R})=\{ S\in GL_{2k}(\N{R}) | S^T {\cal J} S = \pm {\cal J} \},
\end{align}
given in eq.~\eqref{Def:Pk}
is a Lie group:
The associative law and the existence of the identity
  follow from $GL_{2k}(\N{R})$ (the set of all invertible, real $2k\times 2k$ matrices) being a group.
   Define 
\begin{align}
	P^-(k,\N{R})=\{ S\in GL_{2k}(\N{R}) | S^T {\cal J} S = - {\cal J} \}.
\end{align}
The other component of $P(k,\N{R})$ (what we could call $P^+(k,\N{R})$) is $Sp(k,\N{R})$.
Then, if $S^-\in P^-$ and $S^+\in Sp(k,\N{R})$, then we easily see by the definition that 
\begin{align}
S^- S^+,S^+ S^- &\in P^-(k,\N{R}), \nn \\
	S^+_1 S^+_2, S^-_1 S^-_2 &\in Sp(k,\N{R}).
\end{align}
 So the set $P(k,\N{R})$ is closed under group multiplication. This set also includes the inverse of each element. We only have to show this for elements $S\in X^-$, since we already know $Sp(k,\N{R})$ is a Lie group.
 Let $S^T{\cal J} S=- {\cal J}$. Then 
\begin{align}
	(S^T)^{-1}S^T{\cal J} S S^{-1}=(S^T)^{-1}(- {\cal J})S^{-1},
\end{align}
and since we generally have that $ (A^T)^{-1}= (A^{-1})^T$,
\begin{align}
	-{\cal J} = (S^{-1})^T {\cal J} S^{-1},
\end{align}
 so $S^{-1}\in P^-$ too (still, $P^-$ is not a group considered isolated, since it is not closed under group multiplication, and does not include the identity).
 
 We have now derived that $P(k,\N{R})$ is a group. To prove it is a Lie group, we must prove that it is a (topologically) closed subset of $GL_{2k}(\N{R})$:
$f(A)=A^T{\cal J}A$ is a continuous map, the set $\{\pm {\cal J} \}$ is closed in
$GL_{2k}(\N{R})$, and hence $P(k,\N{R})=f^{-1}[\{\pm {\cal J} \}] $ is closed in $GL_{2k}(\N{R})$. 

\paragraph{The determinant of $P^-(k,\N{R})$}
\label{sec:TheDeterminantOfPKNR}

We will now show that the determinant of the matrices in the set $P^-(k,\N{R})$, consisting of the real matrices with the property $S^T \mc{J} S = -\mc{J}$, is $(-1)^k$:

First, we claim the set $P^-(k,\N{R})$ is given by
\begin{align}
	P^-(k,\N{R})=Sp(k,\N{R})\,C =C\, Sp(k,\N{R}),
\end{align}
  with $C$ defined in eq.~\eqref{E:calC}.
  This is so because if $S' \in P^-(k,\N{R})$, then $S'C\in Sp(k,\N{R})$ since
\begin{align}
	(S'C)^T\mc{J}(S'C)= C^T(-\mc{J})C=\mc{J},
\end{align}
  and then $S' = S C$ for $S=S'C \in Sp(k,\N{R})$, since $C^2=I$.
  Similarly with $C\, Sp(k,\N{R})$.
  
On the other hand, if $S \in Sp(k,\N{R})$, then 
\begin{align}
	(SC)^T\mc{J}(SC)=C^T \mc{J} C =-\mc{J},
\end{align}
 so then $S C\in P^-(k,\N{R})$. Similarly, $C S \in P^-(k,\N{R})$.

Now we can evaluate the determinant of an arbitrary element in $S'\in P^-(k,\N{R})$.
Since $S' =S C$ for an element $S\in Sp(k,\N{R})$,
\begin{align}\label{E:detS'}
	\det(S')= \det(S)\det(C)= \det(C),
\end{align}
since all matrices in $Sp(k,\N{R})$ have determinant 1 \cite{Baker}.
The determinant of an $n \times n$ matrix $A$ can be written (sum over repeated indices)
\begin{align}
\det(A) =\epsilon^{i_1,\ldots, i_n} A_{1,i_1}\cdots A_{n,i_n}
\end{align}
 (the Leibniz formula). Then there is only one non-zero term in this
sum for the matrix $C$, so the determinant is given by (no sum over $k$)
\begin{align}\label{E:detR}
	\det(C) &= \epsilon^{1,2,\ldots,2k} C_{1,1}\, C_{2,2}
\cdots C_{2k,\,2k}  = 1^k(-1)^k=(-1)^k.
\end{align}
 Hence by eqs.~\eqref{E:detS'} and \eqref{E:detR}, the matrices of $P^-(k,\N{R})$ have determinant $(-1)^k$. 
  
\paragraph{$Sp(k,\N{R})$ and $P^-(k,\N{R})$ are not connected}
\label{sec:SpKNRAndPKNRAreNotConnected}
We want to show that $Sp(k,\N{R})$ and $P^-(k,\N{R})$ are two components of $P(k,\N{R})$, i.e.~they 
are not connected. Connected means the same as path connected for Lie groups.
Assume that the two components are connected. Then there has to be a continuous path between
e.g.~$I\in Sp(k,\N{R})$ and $C\in P^-(k,\N{R})$. Let $X(t)$ be such a path, i.e.~$X(0)=I$ and
 $X(1)=C$, where $X(t)$ is continuous. Consider the supremum
\begin{align}
	t_0= \sup \{t\,| \, X^T(t) \mc{J} X(t) = +\mc{J} \}.
\end{align}
 We know that $X(1)^T \mc{J} X(1) = -\mc{J}$.
 Moreover, consider the function 
\begin{align}
	f(t)=\det (X^T(t) \mc{J} X(t) +\mc{J}),
\end{align}
which is continuous for continuous functions $X(t)$, since the determinant, matrix addition, multiplication and transposition are continuous. But $f(t)$ is discontinuous for $t=t_0$, since there in any open interval containing $t_0$ will be values $t$ where $f(t)=0$ and other values where $f(t)=\det(2 \mc{J})=2^{2k}$, by definition of $t_0$. Hence our assumption that $X(t)$ is continuous must be wrong,
and hence the sets $Sp(k,\N{R})$ and $P^-(k,\N{R})$ are not connected.

\section{The map $\rho$}
\label{sec:TheMapRho}
We introduce a map $\rho$  which
lets us easily translate between real and complex formulations of the kinetic terms
we are studying. The map $\rho$ preserves both matrix multiplication, addition and the identity.\footnote{$\rho$ is an injective ring homomorphism \cite{Baker}. On the other hand, the inclusion $\rho[U(2)] \subset SO(4)$ shows that $\rho$ does not preserve the determinant, even though it is a ring (or group) isomorphism on its image.}
We define $\rho$ as a function from $M_k(\N{C})$, the set of all $k \times k$ complex matrices, to $M_{2k}(\N{R})$, the set of all $2k \times 2k$ real matrices by
\begin{align}\label{Def:rho}
\rho (X) =
\begin{pmatrix}
 \text{Re}(X) & -\text{Im}(X) \\
 \text{Im}(X) & \text{Re}(X)
\end{pmatrix}.
\end{align}
With $U$ a Lie group,
$\rho$ is a Lie group isomorphism from $U\subset M_k(\N{C})$ to $\rho[U]$.

Now we want to show that the definition of $\rho$ can be extended to vectors so that it preserves products of complex vectors and matrices: Let $v$ be a complex $k\times 1$ vector, let $v=v_R+i v_I$, with $v_R,v_I$ real and define
\begin{align}\label{Def:rhov}
	\rho(v)\equiv \begin{pmatrix}
 \text{Re}(v)  \\
 \text{Im}(v)
\end{pmatrix} = 
\begin{pmatrix}
  v_R  \\
  v_I
\end{pmatrix},
\end{align}
and
\begin{align}\label{Def:rhovdag}
	\rho(v^\dag)\equiv \begin{pmatrix}
 \text{Re}(v^\dag),  &
 -\text{Im}(v^\dag)
\end{pmatrix} = 
\begin{pmatrix}
  v_R^T,  &
  v_I^T
\end{pmatrix}.
\end{align}

Moreover, let $A$ be a complex $k\times k$ matrix and let
 $A=(A_R+iA_I)$, with $A_R, A_I$ real, then
\begin{align}\label{E:rhoAV}
	\rho(A v) =  \rho(A)\rho (v),
\end{align}
 since $\rho(A v) =$ {\scriptsize $\begin{pmatrix} (Av)_R \\ (Av)_I \end{pmatrix}
 = \begin{pmatrix} A_R& - A_I \\ A_I & A_R \end{pmatrix} \begin{pmatrix} v_R \\ v_I \end{pmatrix}$} $= \rho(A) \rho(v)$.
 Furthermore, let $u, v$ be complex $k\times 1$ vectors, then
\begin{align}\label{E:rhoudagAV}
	\text{Re}(u^\dag A v)= \rho(u^\dag) \rho(A) \rho (v),
\end{align}
  since $\text{Re}(u^\dag A v)= \text{Re}[(u^T_R-i u^T_I)(A_R+i A_I)(v_R+i v_I)] =$ {\scriptsize$\begin{pmatrix} u_R^T & u_I^T \end{pmatrix} \begin{pmatrix} A_R& - A_I \\ A_I & A_R \end{pmatrix} \begin{pmatrix} v_R \\ v_I \end{pmatrix}$} $=\rho(u^\dag) \rho(A) \rho (v)$.
Then, 
\begin{align}\label{E:rhoreal}
	u^\dag A v +v^\dag A^\dag u = \rho(u^\dag) \rho(A) \rho (v) + \rho(v^\dag) \rho(A^\dag) \rho (u),
\end{align}
since the left hand side of eq.~\eqref{E:rhoreal} equals its real part.

\end{document}